\documentclass[review]{elsarticle}

\usepackage{lineno,hyperref}
\usepackage{amsmath,epsfig}
\usepackage[ruled,vlined]{algorithm2e}
\modulolinenumbers[5]
\newcommand\norm[1]{\left\lVert#1\right\rVert}
\newcommand\cappednorm{\ell_{p}}
\journal{JVCI}

\begin{document}

\begin{frontmatter}

\title{From Photo Streams to Evolving Situations}
\author[add1]{Mengfan Tang\corref{cor1}}
\ead{mengfant@uci.edu}
\author[add2]{Feiping Nie}
\ead{feipingnie@gmail.com}
\author[add3]{Siripen Pongpaichet}
\ead{siripen.pon@mahidol.ac.th}
\author[add1]{Ramesh Jain}
\ead{jain@ics.uci.edu}
\cortext[cor1]{Corresponding author}
\address[add1]{Department of Computer Science, University of California, Irvine, USA}
\address[add2]{School of Computer Science and Center for OPTical IMagery Analysis and Learning (OPTIMAL), Northwestern Polytechnical University, China}
\address[add3]{Faculty of Information and Communication Technology, Mahidol University, Thailand}




\begin{abstract}
Photos are becoming spontaneous, objective, and universal sources of information. This paper develops evolving situation recognition using photo streams coming from disparate sources combined with the advances of deep learning. Using visual concepts in photos together with space and time information, we formulate the situation detection into a semi-supervised learning framework and propose new graph-based models to solve the problem. To extend the method for unknown situations, we introduce a soft label method which enables the traditional semi-supervised learning framework to accurately predict predefined labels as well as effectively form new clusters. To overcome the noisy data which degrades graph quality, leading to poor recognition results, we take advantage of two kinds of noise-robust norms which can eliminate the adverse effects of outliers in visual concepts and improve the accuracy of situation recognition. Finally, we demonstrate the idea and the effectiveness of the proposed model on Yahoo Flickr Creative Commons 100 Million. 
\end{abstract}


\end{frontmatter}


\section{Introduction}

In today's data-rich environment, big data is readily available from many open sources, proprietary sources, IoTs, and databases. The act of leveraging a human mobility enabled sensing of the environment is referred to as ``Participatory Sensing”.  Increasingly participatory citizen sensing and crowdsourcing are playing  more important roles in understanding current trends and evolving situations. Among diverse  participatory sensing data, photos are one of the biggest data sources. Every minute, millions of photo are uploaded by people to social media. Photos have been used for emerging applications such as event recognition \cite{wang2015tweeting}, trend analysis \cite{jin2010wisdom}, and cultural dynamics \cite{FM4711}.  Photos provide information without a language.

Deep learning frameworks have been successfully used in video and image analysis. Visual concepts can be accurately recognized. This technology of concept detection is commercialized by a company called ``Clarifai'' to provide service in solving real-world problems for businesses and developers for photo understanding.  Accurate concept detector frameworks provide informative and objective multimedia micro-reports. It is a good time to explore use of geo-tagged photos for situation recognition. Different from event detection which is based on data at one particular space and time, situation is defined as the perception of elements in the environment within a volume of time and space, the comprehension of their meaning, and the projection of their status in the near future \cite{endsley1988situation}.
%
%
%
%
We use photos as the information source for radical improvement in the quality of participatory sensing data.  Photo data has associated geo-spatial and other useful information. ``Situation" is characterized by its space-time-theme nature. With information of space and time, the objective and factual nature of photo make it the best resources of situation recognition \cite{siripen2016icmr}. For example, when a situation occurs, people take photos related to the situation, which enables detection of the situation occurrence promptly, simply by observing the photo.   Social situations and trends are usually detected by the concurrences of visual concepts. For example, ``Olympics Games" is always associated with concepts of people, sport, bar, and stadium, etc, evolving in a certain pattern. The event model of ``Olympics Games" can be defined as a bag of these visual concepts. Detection of situations is then transformed to a photo clustering problem, in which each cluster represents a situation \cite{siripen2016icmr}.

Photo clustering assigns photos to groups which share the same semantic concepts of the contents. Many traditional methods, such as K-means, support vector machine, and spectral clustering, can be used to infer photo's label using labeled photo and unlabeled photo. In particular, by using labeled and unlabeled data together, semi-supervised learning can assign labels to the remaining photos by assuming that neighboring data points are likely to have the same label. However, traditional semi-supervised learning frameworks require at least one data sample for each label and lack abilities to discover new situations. 
\begin{figure*}[h]
\centering
\includegraphics[width=1\linewidth]{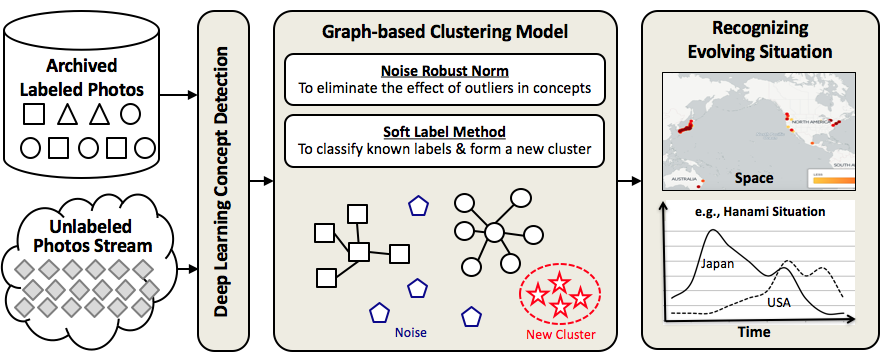}
\caption{Semi-supervised learning Framework for Evolving Situation Recognition under the Condition of Unknown-Labeled Data} \label{figure:flowchart}
\end{figure*}

Graph-based semi-supervised learning methods have been the state-of-the-art in photo annotation and photo understanding \cite{tang2011image}. \cite{zhu2003semi} proposed Harmonic Energy Minimization and use Gaussian fields to propagate label information to unlabeled data. As the proposed method is one kind of random walks, the output can be interpreted as the probabilities of classifying the data points to given labeled clusters. The classification heavily depends on the labeled data which makes it sensitive to the noise in labeled data. Some of the existing graph-based models use a quadratic form of graph embedding. However, a major drawback is the sensitivity of results to outliers. A robust graph-based learning method can overcome this drawback through the user of noise-robust norms. Among these methods, models using $\ell_{0}$-norm have demonstrated effective performance \cite{yang2011robust}. Unfortunately, computational expense increases when subset selection problems. To reduce the computational burden, the $\ell_{0}$-norm has been replaced by some relaxations, such as $\ell_{1}$-norm.

In this paper, we combine the advantages of noise-resistant norms and the soft label methods and propose new graph-based models. This model learns an efficient graph embedding by utilizing $\ell_{1}$-norm and capped $\ell_{p}$-norm to remove data outliers. The soft label method empowers semi-supervised learning framework to accurately predict labels as well as discover new clusters. Furthermore, efficient iterative algorithms are adopted to solve the proposed optimization problem. The proposed framework provides a powerful tool to investigate situation recognition under the condition of noisy and unknown-labeled data. To verify the effectiveness and efficiency of the proposed model, we apply it to Yahoo Flickr Creative Commons data. 

\section{The Proposed Framework}
\subsection{Problem Formulation and Motivation}
Suppose we have millions of photos uploaded to the Web every day. Can we use the photos to observe real-world situations? ``Situation" is defined on space, time and concept information. For example, when a public situation occurs, people take photos related to this situation. People are acting as sensors enabling detection of situation occurrence promptly. We develop a framework that treats photos as micro-reports to detect evolving situations. The framework makes use of visual concepts, space, and time information and includes three components: data collection, clustering, and filtering by space and time.  An overview of the proposed framework is shown in Fig. 1. 
The framework first collects historical photos and their labeled situations. For each photo, deep learning concept detector is applied to get visual concepts. We do not aim at developing new concept detection methods or try to improve the existing concept detector. We use these concepts as features to detect situations. At the data collection stage, photos are associated with visual concepts and situation labels. It is commonly known that the number of labeled data is far less than unlabeled data on the web. Semi-supervised learning methods can be naturally adopted to target the problem: recognizing new photo labels. However, traditional semi-supervised learning methods can not be directly applied because they require that photos must belong to at least one predefined cluster. In real world applications, unknown or new situations may exist in unseen photos. One the other hand, noise may also exist in the new data. Thus, in the stage of clustering , we need to handle both the new situation problem and the noise problem. We introduce a soft label method which can effectively form new clusters as well as accurately predict known labels. We incorporate noise robust norms to eliminate the adverse effects of outliers in visual concepts and thus improve the accuracy of situation recognition. At the end of the second stage, photos are labeled by situations. At the last stage, we use space and time information to further understand when and where these situations happen. For example, given photos of the situation ``Olympic Games", time information can be used to understand if it is a ``Winter Olympic Games" or a ``Summer Olympic Games". Another example from space perspective is ``holi".  ``Holi" is an original Indian festival. Gradually it has been celebrated in the United States and Europe. The proposed framework can show how the ``holi" trend evolving at different locations. 

\subsection{Semi-supervised Learning Method for Unknown Labels}
Given $n$ photos $\{x_{1},\cdots,x_{m},x_{m+1},\cdots,x_{n}\}$ and the labeled situation set $U = \{1,\cdots,u\}$, $\{x_{1},\cdots,x_{m}\}$ are labeled with known situations $y\in U$ and the remaining photos $\{x_{m+1},\cdots,x_{n}\}$ are not labeled.  The goal is to use visual concepts of the photos as features to predict the labels of unlabeled photos and discovery a new situation if there is in the photos. Predicting the labels of unlabeled photos using both labeled and unlabeled data is a semi-supervised learning problem. To enable the traditional semi-supervised learning framework to new situation discovery, we introduce one more label variable denoted as $u+1$, where $\hat{U} = \{1,\cdots,u+1\}$. This simple setting will solve the new label discovery problem which traditional semi-supervised learning framework cannot solve.

Consider a connected graph of photos, $G = (V,E)$. Nodes in $V$ correspond to the $n$ data points. Nodes in $M =(1,2,\cdots,m)$ are situation-labeled photos. Nodes in $U = (m+1,\cdots,m+u)$ are unknown-situation label photos. The edge $E$ in the graph is described in the similarity matrix $W\in R^{n\times n}$, where $w_{ij}$ is the similarity measurement of a pair of vertices, $x_{i}$ and $x_{j}$.
For the similarity matrix $w_{ij}$, one of the examples using Gaussian function is,
\begin{equation*}
w_{ij}= 
\begin{cases}
    \exp(-\sum_{z = 1}^p \frac{(x_{iz}-x_{jz})^2}{\sigma_{z}^2})& j\in \mathcal{N}_{i} \enspace \text{or} \enspace i\in \mathcal{N}_{j}\\
    0,              & \text{otherwise}
\end{cases}
\end{equation*}
where $x_{i} = (x_{i1},\cdots,x_{iz})$, $x$ has $p$ dimensions of features. $\mathcal{N}_{i}$ is a set of indices of $x_{i}$'s neighbors, $\sigma_{z} (z = 1,\cdots,p)$ are parameters associated with features.

Given a similarity matrix $W$,  The key idea of the graph based method is that the nodes connected by a large weight in $W$ in the graph have similar values. In other words, observations $y_{1},\cdots,y_{n}$ change smoothly on the graph. Based on the smoothness assumption, the task is to assign labels to nodes $U$.  \cite{nie2010general} proposed a general graph-based semi-supervised learning method for new class discovery.

\begin{equation}\label{semi}
\min_{\substack{F}}\sum_{i,j}\hat{w}_{ij} \norm{f_{i}-f_{j}}_{2}^2+\sum_{i=1}^{n}u_{i}\hat{d_{i}} \norm{f_{i}-y_{i}}_{2}^2,
\end{equation}
where the normalized weights $\hat{w_{ij}}$can be computed by $\hat{w}_{ij} = w_{ij}/\sqrt{d_{i}d_{j}}$, $d_{i} = \sum_{j}w_{ij}$. 
By optimizing  Eq. (\ref{semi}),  we can get the soft label matrix $F \in \mathcal{R}^{n\times (c+1)}$.  Then the  label of $x_{i}$ can be calculated as ,
\begin{equation*}
y_{i} = \arg\max_{j\leq c+1}F_{ij}.
\end{equation*}
This model has two terms. The first term plays a role as a regularization, which controls the smoothness of the predicted labels on the graph. The second term determines the degree of label matching between the predicted labels and initial labels. Two parameters $u$ and $d$ are used to balance the trade-off between these two terms.

\subsection{Noise Robust Semi-Supervised Learning Methods for Unknown Labels}

Traditional methods such as (\ref{semi}) are built on $\ell_{2}$-norm of graph embedding. The quadratic form makes these methods sensitive to noise or outliers. To overcome the noisy data which degrades graph quality, leading to poor recognition results, we use noise robust norms which can eliminate the adverse effects of outliers in visual concepts and improve the accuracy of situation recognition. In particular, we propose one $\ell_{1}$-norm method and one capped norm method. 
Capped norm based loss function has been used for various purposes. For example, capped norm has been used for unsupervised photo clustering \cite{tang2016capped}. Here, we use $\cappednorm$-norm as a robust and stable loss function to resist outliers.
We use this property in projecting data into a manifold where the similarity of data points is adjusted without input from outliers. If the distance of a label vector is large, the corresponding similarity value is not updated.

\begin{equation}\label{l1semi}
\min_{\substack{F}}\sum_{i,j}\hat{w}_{ij} \norm{f_{i}-f_{j}}_{2}+\sum_{i=1}^{n}u_{i}\hat{d_{i}} \norm{f_{i}-y_{i}}_{2}^2.
\end{equation}

Capped norm based loss function of a $c$-dimensional vector $u\in\mathcal{R}^{1 \times c}$ is $\min(\norm{u}_{2}^{p},\theta)$, where $\theta$ is a parameter. The value of this loss function is $\norm{u}_{2}^{p}$, if $\norm{u}_{2}^{p}$ is smaller than $\theta$, and is $\theta$, otherwise. This loss function is more robust to outliers than $\ell_{1}$-norm because it has threshold $\theta$ for outliers.

The proposed Capped $\cappednorm$-Norm Method is expressed as,
\begin{equation}
\label{cappednorm}
\min_{\substack{F}}\sum_{i,j}\hat{w}_{ij} \min(\norm{f_{i}-f_{j}}_{2}^p,\theta)+\sum_{i=1}^{n}u_{i}\hat{d_{i}} \norm{f_{i}-y_{i}}_{2}^2,
\end{equation}
where $w_{ij}$ is the similarity measurement in graph between $x_{i}$ and $x_{j}$, $0<p\leq 2$ and $\theta$ is a parameter.

The above formulation benefits from the input control of input data. The clustering results are dependent on the quality of input data graph. Most of the time, they are sensitive to the particular graph construction methods. We overcome this problem from two perspectives which are graph initialization, and graph similarity adaptation.  We will introduce a method to give a ``good" initialization.  For the graph similarity adaptation, if the distance of label vector is large, then the corresponding similarity value is not updated because of the capped norm. 
By optimizing  Eq. (\ref{l1semi}) and (\ref{cappednorm}),  we can get the soft label matrix $F \in \mathcal{R}^{n\times (c+1)}$.  Similar to Eq.(\ref{semi}), the  label of $x_{i}$ can be calculated as ,
\begin{equation*}
y_{i} = \arg\max_{j\leq c+1}F_{ij}.
\end{equation*}
\subsection{Optimization Algorithm}
In this subsection, we introduce the optimization algorithm to solve problem (\ref{l1semi}) and (\ref{cappednorm}). 
To optimize the objective function easily, we rewrite \ref{l1semi} into a matrix form,
\begin{equation*}\label{l1semimatrix}
\min_{\substack{F}}\sum_{i,j}\hat{w}_{ij} \norm{f_{i}-f_{j}}_{2}+Tr(F-Y)^TU\hat{D}(F-Y),
\end{equation*}
where $\hat{D}$ is a diagonal matrix with diagonal entries as $\hat{D_{ii}} = \sum_{j} \hat{W}_{ij},\forall i$.
Taking derivative of \ref{l1semimatrix} with respect to $F$, and setting the derivative to zero, we have,
\begin{equation*}\label{kktl1}
\bar{L}F+U\hat{D}(F-Y) = 0,
\end{equation*}
where $\bar{L}$ is the Laplacian matrix of $\bar{W}$, and the $ij$-$th$ element of $\bar{W}$ is defined by
\begin{equation*}
\bar{W}_{ij} = \frac{W_{ij}}{2\norm{f_{i}-f_{j}}_{2}}
\end{equation*}

Thus, 
\begin{equation*}
F = (\bar{L}+U\hat{D})^{-1}U\hat{D}Y
\end{equation*}

It's noted that $\bar{L}$ is dependent on $F$, we propose an iterative algorithm to obtain the solution.

In the objective function, similarity matrix $W\in \mathcal{R}^{n\times n}$ is required, and the structure of the graph plays important roles in the performance of the graph-based clustering. We use a method \cite{nie2016constrained} to generate initial graph for the proposed model. This method has only one integer parameter: the number of neighbors, which is easier to tune.
\begin{equation*}
\min_{w_{i}^T\mathbf{1}=1,w_{i}\geq 0,w_{ii}=0}\sum_{j=1}^{n}\norm{x_{i}-x_{j}}_{2}^2w_{ij}+\lambda\sum_{j=1}^{n}w_{ij}^2.
\end{equation*}
where $w_{i}$ is the $i$-th row of similarity matrix $W$.
Denote $e_{ij} = \norm{x_{i}-x_{j}}_{2}^2$, and use Lagrangian method, the optimal similarities can be obtained,
\begin{equation}
\label{graphconstruct}
\hat{w_{ij}} =  
\begin{cases}
\frac{e_{i,m+1}-e_{ij}}{ke_{i,k+1}-\sum_{m=1}^{k}e_{ik}} & j\leq k\\
0 & j>k\\
\end{cases}
\end{equation}
where $k$ is the number of neighbors. Because of the simplicity of computation of $\hat{w_{ij}}$, compared to Gaussian functions, it is fitting into a large-scale graph construction.

\begin{algorithm}[h]
Initialize $W_{ij}$;\\
Construct graph for $w_{ij} \in W$ using Equation (\ref{graphconstruct}).\\
\Repeat{converge}{
Calculate $\bar{L_{t}} = \bar{D_{t}}-\bar{W_{t}}$,\\ 
where $\bar{(W_{t})}_{ij} = \frac{W_{ij}}{2\norm{(f_{t})_{i}-(f_{t})_{j}}_{2}}$\\
Calculate $F_{t+1} = (\bar{L_{t}}+U\hat{D})^{-1}U\hat{D}Y$\\
t = t+1;\\
}
Assign labels to $x_{i}$, $y_{i} = \arg\max_{j\leq c+1}F_{ij}$
\caption{{\bf Semi-supervised $\ell_{1}$-norm Method} \label{Algorithm}}
\end{algorithm}

To optimize the objective function in the Problem (\ref{cappednorm}) easily, we rewrite it into a matrix form,

\begin{equation*}\label{cappednormmatrix}
\sum_{i,j}\hat{w}_{ij} \min(\norm{f_{i}-f_{j}}_{2}^p,\theta)++Tr(F-Y)^TU\hat{D}(F-Y),
\end{equation*}
Because the proposed model is a weighted sum of a concave function, it can be solved in a general re-weighted optimization framework \cite{nie2014optimal}. 

The general re-weighted optimization problem is,
\begin{equation}
\label{general}
\min_{x\in \mathcal{C}} f(x)+\sum_{i} h_{i}(g_{i}(x)),
\end{equation}
where $h_{i}(x)$ is an arbitrary concave function with the domain of $g_{i}(x)$.
This general problem can be solved by Algorithm $1$.
\begin{algorithm}[h]
Initialize $X\in \mathcal{C}$;\\
\Repeat{converge}{Calculate $D_{i} = h^{\prime}_{i}(g_{i}(x))$\\
Solve the following problem $\min_{x\in\mathcal{C}}f(x)+\sum_{i}Tr(D_{i}^{T}g_{i}(x))$\\}
\caption{{\bf Solving the Problem (\ref{general})} \label{Algorithm}}
\end{algorithm}

For problem ($\ref{cappednorm}$), denote $h(x) = \min(x^{\frac{p}{2}},\theta)$ and $x = \norm{f_{i}-f_{j}}_2^2$. Because $h(x)$ is a concave function with respect to $x$, the supergradient of $h(x)$ can be obtained.\\
\begin{equation*}
h^{\prime}(x) = 
\begin{cases}
\frac{p}{2}x^{\frac{p-2}{2}}& x^{\frac{p}{2}}\leq \theta\\
0& \text{otherwise}
\end{cases}
\end{equation*}

Problem ($\ref{cappednorm}$) can be solved using Algorithm $1$, because $h(x)$ is concave. The first part of the proposed model can be written in the following form,
\begin{equation*}
\begin{split}
\min_{F^TF=I}\sum_{i,j}w_{ij}h^{\prime}(x)x & =\min_{F^TF=I}\sum_{i,j}w_{ij}s_{ij}\norm{f_{i}-f_{j}}_2^2\\
&= \min_{F^TF=I}\sum_{i,j}\tilde{s}_{ij}\norm{f_{i}-f_{j}}_2^2\\
&=\min_{F^TF=I} Tr(F^TL_{\tilde{s}}F),
\end{split}
\end{equation*}
where $\tilde{s_{ij}} = w_{ij}s_{ij}$.

Fixing $\tilde{s_{ij}}$ , the objective function becomes,
\begin{equation}
\label{reducedfunction}
\min_{F}Tr(F^TL_{\tilde{s}}F)+Tr(F-Y)^TU\hat{D}(F-Y),
\end{equation}

Taking  derivative of \ref{reducedfunction} with respect to $F$, and setting the derivative to zero, we have,
\begin{equation*}\label{kkt}
L_{\tilde{s}}F+U\hat{D}(F-Y) = 0.
\end{equation*}
Thus, 
\begin{equation*}
F = (L_{\tilde{s}}+U\hat{D})^{-1}U\hat{D}Y
\end{equation*}
We propose an iterative algorithm to obtain the solution. The algorithm is guaranteed to converge because it is an application of a general optimization framework.

We apply an iterative algorithm to solve the optimization
problem ($4$). One of the most important aspects of iterative
algorithm is convergence. The objective function belongs to the general optimization problem. Naturally, it is theoretically proven to converge \cite{nie2014optimal}.

According to Algorithm $2$, we have the following Algorithm $3$ to solve the Problem (\ref{cappednorm}).
\begin{algorithm}[h]
Initialize $s_{ij}$ and set $s_{ij}=1$;\\
Construct graph for $w_{ij} \in W$ using Equation (\ref{graphconstruct}).\\
\Repeat{converge}{
Calculate $\tilde{w}$ by $\tilde{w} = w_{ij}s_{ij}$\\
Update $F$ by $F = (L_{\tilde{s}}+U\hat{D})^{-1}U\hat{D}Y$;\\
Calculate $s_{ij} = \begin{cases}
\frac{p}{2}\norm{f_{i}-f_{j}}_{2}^{p-2}& \norm{f_{i}-f_{j}}^{p}_{2}\leq \theta\\
0& \text{otherwise}
\end{cases}$
}
Assign labels to $x_{i}$, $y_{i} = \arg\max_{j\leq c+1}F_{ij}$
\caption{{\bf  Semi-supervised $\cappednorm$-norm Method} \label{Algorithm}}
\end{algorithm}

\section{Experiments}
\subsection{Data Pre-processing}

Yahoo Flickr Creative Commons 100M (YFCC100M) dataset \cite{thomee2016yfcc100m} is used to demonstrate our idea.  As we described in previous sections, deep-learning approaches from Clarifai is used to detect 1,570 concepts in all the sampled photos. We do not aim at developing new concept detection methods or try to improve the existing concept detector. We use these concepts as features to detect situations.  The top-500 concepts are used as features because of a long-tailed distribution of concepts.  A list of top-30 concepts is shown as {`people',
 `nature',
 `indoor',
 `sport',
 `landscape',
 `plant',
 `architecture',
 `music',
 `performer',
 `tree',
 `demonstrator',
 `vehicle',
 `building',
 `concert',
 `cherry blossom',
 `water',
 `musician',
 `blossom',
 `crowd',
 `sakura',
 `outfit',
 `ensemble',
 `text',
 `stage',
 `slope',
 `riverbank',
 `animal',
 `lake',
 `road', and
 `sign'}.

We perform two experiments to verify the idea of situation recognition and the proposed models. The first one is to prove that the proposed noise-robust norms are effective in reducing the adverse function of outliers. The second experiment is to apply the verified models to YFCC100M data and show the evolving situation recognition. 
\begin{table}
\centering
\caption{Accuracy(in percentage) for label-known data and label-unknown data: GSS is Semi-supervised Learning Method for Unknown Labels Model (Eq. ($\ref{semi}$)), SSL is $\ell_{1}$-norm Semi-supervised Learning Method (Eq. ($\ref{l1semi}$)), SSC is $\cappednorm$-norm Semi-supervised Learning Method (Eq. ($\ref{cappednorm}$))}
\vspace{1\baselineskip}
\begin{tabular}{c|c|c}
\cline{1-3}
\textbf{Method}   & \textbf{Label-known Data}  & \textbf{Label-unknown Data}\\ \cline{1-3}
GSS & 92.80 & 91.40\\
SSL & 93.64 & 97.40\\ 
SSC & 96.06 & 95.10 \\ \cline{1-3}
\end{tabular}
\label{taresults}
\end{table}

\begin{figure}[h]
\centering
\includegraphics[width=0.45\linewidth]{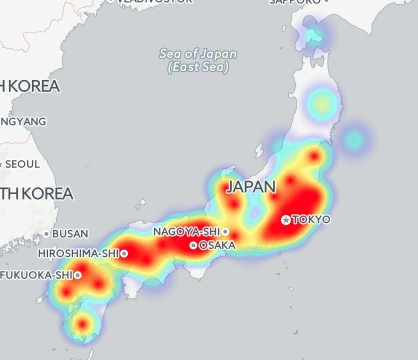}
\includegraphics[width=0.5\linewidth]{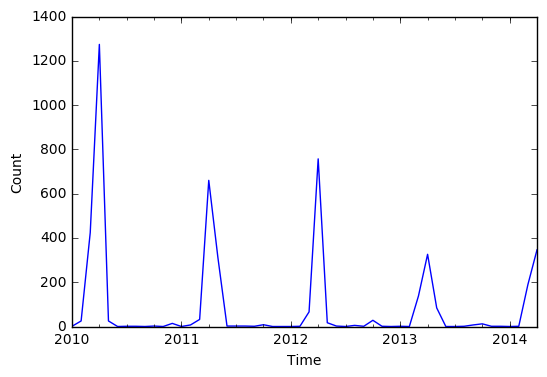}
\caption{The Evolving Situation of ``Hanami"} \label{hanami}
\end{figure}
\begin{figure}[h]
\centering
\includegraphics[width=0.45\linewidth]{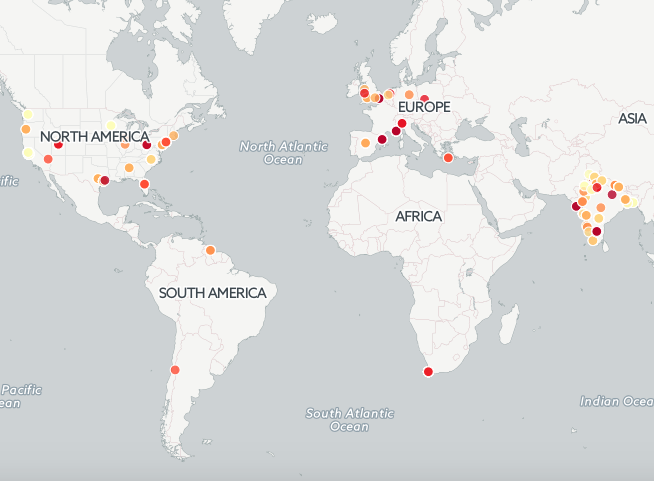}
\includegraphics[width=0.5\linewidth]{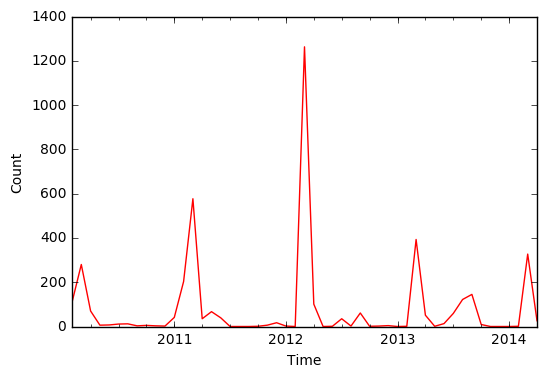}
\caption{The Situation of ``Holi" in Space and Time} \label{hanami}
\end{figure}
\subsection{Semi-supervised Clustering with Unknown Labels}
We use six situations to demonstrate the idea and the proposed framework. The six situations include ``hanami", ``Olympic Games",``protest", ``flood", ``blues fest", and ``holi". We use the users' tags as situation labels.  We randomly select 500 photos for each situation before the year of 2010 as the labeled data, then we randomly select 3000 photos after the year of 2010 as testing data. Among the unlabeled data, 500 photos are not in the categories of  predefined situations as noise. We use 3000 labeled photos and 3000 unlabeled photos. We present experimental study on GSS \cite{nie2010general}, Semi-supervised $\ell_{1}$-norm method(SSL), and Semi-supervised $\cappednorm$-norm method(SSC) for situation detection. There are parameters $u_{i}$ in the proposed methods. For fair comparison, we set the compared methods and the proposed methods with the same parameter setting. We use one $u$ for the labeled data and another $u$ for unlabeled data to control the smoothness term and the fitting term for the model. $u = (1,10,20,30,40,50,60,70,80,90,100)$. We also do cross validation to select the parameter combination for SSC from the range $\theta = (0.01,0.1,1,10)$ and $p = (0.5,0.7,1,1.5,1.7,2)$. 
%
%

As Table 1 shows, two proposed models SSL and SSC outperform the $\ell_{2}$-norm based GSS model. For the known labeled data, compared with GSS, SSC improves the clustering accuracy  by $4.3\%$, and SSC improves the accuracy from $92.80\%$ to  $93.64\%$. For the unknown labeled data, both SSL and SSC consistently improved over the benchmark. Methods such as traditional semi-supervised learning methods are not able to assign unknown labels to data, the accuracy of those methods on unknown labeled data is 0. Compared with GSS, the proposed methods SSC and SSL reduce testing error by $6.5\%$ and $4.0\%$, respectively, which further proves the effectiveness of our methods in  discovering unknown labels. 
%
%
%
%
%
%
%

\subsection{Evolving Situation Detection}
%
%
In order to detect the evolving situations, we further analyze space and time context of the labeled photo outputs from our clustering method and apply the proposed framework to the data after the year 2011 to detect situations. 
Fig. 2 shows that we can successfully detect the situation of ``hanami'' in Japan. Tokyo, Osaka, and Hiroshima are the three most popular places for ``hanami''.  It happened in late March and early April every year.  Similarly, we  demonstrate it on detection of ``holi''. ``Holi'' is an Indian festival. People from the United States and Europe have started to celebrate this festival recently. Fig. 3 shows that ``holi'' is  celebrated not only in India but is widely celebrated in the United States and Europe also in March every year.  

\section{Conclusion}
Situation recognition is important in many real world applications ranging from disaster response to politics and social happenings.  With the explosive growth of user-contributed photos and videos, there is a great potential opportunity to detect evolving situations for responding to them.  Combining geo-tagged photo streams and  concept detection,  it is  a  right time to explore situation recognition. We investigate the use of visual concepts for situation recognition and develop new methodologies to tackle some technical challenges lying in noisy data and existing approaches. Situation recognition will be enriched by developing methods to combine visual concepts with physical sensory data. Real-time situation recognition and  predictive  analysis  are important components of many new challenges faciltated by big data.

\section*{References}

\bibliography{mybibfile}

\end{document}